\documentclass[prd,superscriptaddress, nofootinbib]{revtex4}
\usepackage{graphicx}
\usepackage{dcolumn}
\usepackage{bm}
\usepackage{amsmath,amsthm,amssymb}

\begin{document}
%


\title{Late-time tails of a Yang-Mills field\\ on Minkowski and Schwarzschild backgrounds}

\author{Piotr Bizo\'n}
\affiliation{M. Smoluchowski Institute of Physics, Jagiellonian
University, Krak\'ow, Poland} \affiliation{Max-Planck-Institut f\"ur
Gravitationsphysik, Albert-Einstein-Institut, Potsdam, Germany}
\author{Tadeusz Chmaj}
\affiliation{H. Niewodniczanski Institute of Nuclear
   Physics, Polish Academy of Sciences,  Krak\'ow, Poland}
   \affiliation{Cracow University of Technology, Krak\'ow,
    Poland}
\author{Andrzej Rostworowski}
\affiliation{M. Smoluchowski Institute of Physics, Jagiellonian
University, Krak\'ow, Poland}
\date{\today}
\begin{abstract}
We study the late-time behavior of spherically symmetric solutions
of the Yang-Mills equations on Minkowski and Schwarzschild
backgrounds. Using nonlinear perturbation theory we show in both
cases  that solutions having smooth compactly supported initial data
posses tails which decay as $t^{-4}$ at timelike infinity. Moreover,
for small initial data on Minkowski background we derive the
third-order formula for the amplitude of the tail and confirm
numerically its accuracy.

\end{abstract}

\maketitle

\section{Introduction} In a classical paper \cite{em} Eardley and
Moncrief proved that solutions of the Yang-Mills equations on the
$3+1$ Minkowski spacetime starting from smooth initial data remain
smooth for all future times. A different proof allowing for initial
data with only finite energy was given later by Klainerman and
Machedon \cite{km}. Once global regularity was established, the
problem of asymptotic behavior of solutions for $t\rightarrow
\infty$ was studied by many authors \cite{gs, ch1, cpc, shu, sch}
who obtained various decay estimates using different techniques and
assumptions about initial data. In this paper we are concerned with
the simplest possible situation, namely
 spherically
symmetric initial data with compact support. In this case it follows
from the conformal method of Christodoulou that the Yang-Mills
curvature decays as $t^{-4}$ at timelike infinity \cite{ch1}. The
purpose of this paper is threefold. First, we rederive
Christodoulou's result using the nonlinear perturbation theory. The
advantage of our approach lies in its wide applicability; in
contrast to the conformal method which is very powerful (in the
sense of giving sharp decay rates) only for conformally invariant
equations.

Second, we go beyond qualitative decay estimates and give the
third-order  formula for the amplitude of solution which provides a
precise \emph{quantitative} information about the tail. We wish to
point out that although our result depends crucially on spherical
symmetry, the assumption of compact support for initial data is made
for simplicity and can be relaxed by imposing a suitable fall-off
condition  at spatial infinity (which can be implemented via
appropriately weighted norms). However, some kind of localization
condition is necessary in order to avoid a situation where the tail
in time is induced entirely by the tail of initial data at spatial
infinity (due, for instance, to nonzero charge).

Third, we argue that the same tail is present in the scattering of
spherically symmetric Yang-Mills fields off the Schwarzschild black
hole. In this case the global existence of solutions follows from
the work of Chru\'sciel and Shatah \cite{cs} who generalized the
proof of Eardley and Moncrief to arbitrary globally hyperbolic
Lorentzian 4-manifolds. The late-time tail of the Yang-Mills field
on the Schwarzschild background was studied in \cite{cw}, however
the fall-off $t^{-5}$ derived there on the basis of the linear
perturbation analysis is not correct. As we shall see, the error in
\cite{cw} is due to the fact that the late-time tail is \emph{not}
governed by the linearized evolution. At first sight that might seem
odd but upon reflection it is easy to understand. A rough intuitive
explanation is that the tail is a far-field  effect hence the flat
space tail $t^{-4}$ is expected to persist in any asymptotically
flat spacetime as long as the backscattering on the curvature does
not produce a more slowly decaying tail.
 A similar example of the failure of linear perturbation analysis
 was recently observed in the scattering of
skyrmions \cite{bcr}. \section{Minkowski background.} We consider
the Yang-Mills theory with the gauge group $SU(2)$ and assume the
spherically symmetric ansatz for the connection \cite{fm}
\begin{equation}\label{ansatz}
    A=w\, \tau_1 d\theta +(\cot\theta\, \tau_3+w\,\tau_2)\sin\theta
    \,d\phi\,,
\end{equation}
where $w=w(t,r)$ and $\tau_i$ ($i=1,2,3$) are the usual generators
of $su(2)$. The Yang-Mills equations $d\ast F=0$, where
$F=dA+A\wedge A$ is the Yang-Mills curvature,
 reduce then to the  semilinear radial wave
equation
\begin{equation}\label{eqw}
    \ddot w - w''-\frac{1}{r^2} w (1-w^2)=0\,,
\end{equation}
where primes and dots denote derivatives with respect to $r$ and
$t$, respectively.  For our purposes it is convenient to define the
function $f(t,r)=(w(t,r)-1)/r$ and rewrite equation (\ref{eqw}) in
the following form
\begin{equation}\label{ym}
 \mathcal{L} f := \ddot f - f'' - \frac{2}{r} f'+ \frac{2}{r^2}f= -f^3-\frac{3}{r}
 f^2\,.
\end{equation}
Note that $\mathcal{L}$  is the   radial wave operator for the $l=1$
spherical harmonic.

 We consider  the late-time evolution of solutions
of equation (\ref{ym}) for smooth compactly supported initial data
\begin{equation}\label{id}
    f(0,r)=\varepsilon \alpha(r),\qquad \dot f(0,r)=\varepsilon
    \beta(r)\,.
\end{equation}
The prefactor $\varepsilon$ is added for convenience and to
emphasize that our initial data are assumed to be small.  Regularity
at the origin is ensured by the boundary condition $f(t,r)\sim
b(t)r$ for $r\rightarrow 0$. As follows from \cite{gs} such
solutions decay to zero on any compact region of space as
$t\rightarrow\infty$.
To determine  the asymptotic behavior of solutions we define the
perturbative expansion
\begin{equation}\label{pert}
    f=\varepsilon f_1 + \varepsilon^2 f_2+\varepsilon^3 f_3+...\,,
\end{equation}
where $\varepsilon f_1$ satisfies initial data (\ref{id}) and all
$f_n$ with $n>1$ have zero initial data. Substituting the expansion
(\ref{pert}) into equation (\ref{ym}) up to the third order we get
\begin{eqnarray}
  \mathcal{L} f_1 &=& 0\,, \label{f1p}\\
  \mathcal{L} f_2 &=& -\frac{3}{r} f_1^2\,, \label{f2p}\\
  \mathcal{L} f_3 &=& -f_1^3-\frac{6}{r} f_1 f_2\,. \label{f3p}
\end{eqnarray}
We solve these equations recursively.  The first order solution is
given by the general regular solution of the free radial wave
equation for the $l=1$ spherical harmonic
\begin{equation}\label{f1}
    f_1(t,r)=\frac{a'(t-r)+a'(t+r)}{r}+\frac{a(t-r)-a(t+r)}{r^2}\,,
\end{equation}
where the function $a(\xi)$ is determined by the initial data
\begin{equation}\label{a}
    a(\xi)=-\frac{1}{2}\xi\int\limits_{\xi}^{\infty}
    \alpha(s)ds + \frac{1}{4} \int\limits_{\xi}^{\infty} (s^2-\xi^2)
     \beta(s) ds\,.
\end{equation}
 For
compactly supported initial data the function $a(\xi)$ has compact
support as well (note that  the functions $\alpha(s)$ and $\beta(s)$
in (\ref{a}) are odd extensions of initial data to the whole line),
hence $f_1$ has no tail in agreement with Huygens' principle.

 To solve equations for the higher order perturbations we use
   the retarded Green's function of the operator $\mathcal{L}$
\begin{equation}\label{green}
    G(t-t',r,r')=[|r-r'|\leq t-t'\leq r+r'] \, \frac{r^2+{r'}^2-(t-t')^2}{4
    r^2}\,.
\end{equation}
It follows from (\ref{green}) that the solution of the inhomogeneous
equation $\mathcal{L} f = N(t,r)$ with zero initial data has the
form (using null coordinates $u=t'-r', v=t'+r'$) \cite{gs}
\begin{equation}\label{duh2}
    f(t,r)= \frac{1}{8 r^2}
    \int\limits_{|t-r|}^{t+r} dv \int\limits_{-v}^{t-r} K(t,r;u,v)
    N(u,v) du\,,
\end{equation}
where the kernel $K(t,r;u,v)= (v-t)(t-u)+r^2$.
  In the second order, i.e. for equation (\ref{f2p}),
the representation (\ref{duh2}) yields
\begin{equation}\label{eqf2}
    f_2(t,r)= -\frac{3}{4r^2}
    \int\limits_{|t-r|}^{t+r} dv  \int\limits_{-v}^{t-r}
    K(t,r;u,v) \frac{f_1^2(u,v)}{v-u}
    du\,.
\end{equation}
Somewhat surprisingly, Huygens' property is preserved in the second
order. To see this, let us assume that $a(\xi)=0$ for $|\xi|\geq R$.
Then, for $t>r+R$ we may change the order of integration in
(\ref{eqf2}) and rewrite it as (see Fig.~1)
\begin{equation}\label{eqf2p}
    f_2(t,r)= -\frac{3}{r^2} \int\limits_{-R}^{R} du
    \int\limits_{t-r}^{t+r} \frac{(v-t)(t-u)+r^2}{(v-u)^3}
    \left(a'(u)+\frac{2 a(u)}{v-u}\right)^2
    dv\,.
\end{equation}
Performing the inner integral we get
\begin{equation}\label{f2as}
    f_2(t,r)= 8 r
    \int\limits_{-R}^{R}
     \frac{a(u)}{(t-u)^2-r^2}\frac{d}{du}\left(\frac{a(u)}{(t-u)^2-r^2}\right) du
    \,,
\end{equation}
which after integration  gives zero. Thus, $f_2(t,r)$ vanishes
identically for $t>r+R$  and consequently there is no tail up to the
second order.

\begin{figure}[tbh]
\centering
\includegraphics[width=0.4\textwidth]{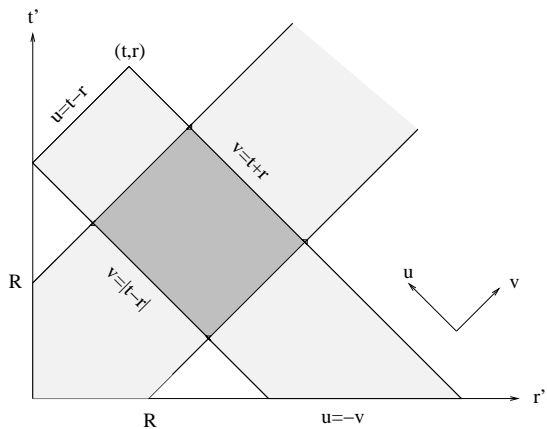}
\caption{\small{{ An illustration of the situation in equations
(\ref{eqf2p}), (\ref{eqf31}), and (\ref{eqf32}). The observation
point is located at $(t,r)$ where $t>r+R$. The integration range is
given by the intersection of the two shaded regions which depict the
domain of dependence of the observation point and the support of the
solution $f_1(t',r')$.}} }\label{fig1}
\end{figure}

 In the third order, i.e. for equation (\ref{f3p}),
the representation (\ref{duh2}) gives $f_3=f^{(1)}_3+f^{(2)}_3$,
where
\begin{eqnarray}
    f^{(1)}_3(t,r)&=& -\frac{1}{8r^2}
    \int\limits_{|t-r|}^{t+r} dv \int\limits_{-v}^{t-r} K(t,r;u,v) f_1^3(u,v)
    du\,,\label{eqf31} \\
    f^{(2)}_3(t,r)&=& -\frac{3}{2r^2}
    \int\limits_{|t-r|}^{t+r} dv \int\limits_{-v}^{t-r} K(t,r;u,v) \frac{
    f_1(u,v) f_2(u,v)}{v-u}
    du\,. \label{eqf32}
\end{eqnarray}
To calculate $f^{(1)}_3(t,r)$ for $t>r+R$, as above we change the
order of integration  and perform the integral over $v$ with the
result (using the abbreviation $z=(t-u)^2-r^2$)
\begin{equation}\label{f3exact}
    f_3^{(1)}(t,r)= 4 r
    \int\limits_{-R}^{R}
     \left(\frac{a(u) a'(u)^2}{z^2} + \frac{4(t-u)  a'(u) a^2(u)}{z^3} + \frac{4((t-u)^2+\tfrac{1}{5}r^2)  a^3(u)}{z^4}\right) du
    \,,
\end{equation}
which has the following asymptotic behavior near timelike infinity
($r=const$ and $t\rightarrow \infty$)
\begin{equation}\label{tail1}
   f^{(1)}_3(t,r) \sim c_1 \, r\, t^{-4}\,,\qquad  c_1=4\int\limits_{-\infty}^{+\infty} a(u)
{a'(u)}^2    du\,.
    \end{equation}
In the formula above we  replaced $R$ by $\infty$ in the limits of
integration to emphasize that the result holds not only for strictly
compactly supported initial data but also for initial data which
fall off sufficiently fast at spatial infinity.

    To calculate the contribution to the tail coming from $f^{(2)}_3(t,r)$
    we need to know both the leading and the subleading terms in the asymptotic expansion of $f_2(u,v)$ near null infinity ($u=const$ and $v\rightarrow\infty$).
     This calculation is
    deferred to the appendix where we show that near null infinity
\begin{equation}\label{f2nul}
 f_2(u,v) =
 \frac{h'(u)}{v-u}+\frac{2h(u)}{(v-u)^2}+\frac{2 g(u)}{(v-u)^2}+\mathcal{O}(v^{-3})\,,
\end{equation}
where $h(u)$ and $g(u)$ are defined by (\ref{h}) and (\ref{g}),
respectively. Note that the first two terms in (\ref{f2nul})
represent the "free" part of the iterate $f_2(t,r)$; as we shall see
in a moment this part does \emph{not} affect the behavior of
$f^{(2)}_3(t,r)$ at timelike infinity.
Substituting (\ref{f2nul}) into (\ref{eqf32}) and proceeding along
the same lines as in the derivation of the expression (\ref{tail1})
we obtain the following asymptotic behavior near timelike infinity
\begin{equation}\label{tail2}
   f^{(2)}_3(t,r) \sim c_2 \, r\, t^{-4}\,,\qquad  c_2= 4
   \int\limits_{-\infty}^{+\infty} \left[\frac{d}{du}(h(u) a(u)) + g(u) a'(u)\right] du = -12
   \int\limits_{-\infty}^{+\infty} a(u) a'(u)^2 du\,,
    \end{equation}
    where the last expression follows from (\ref{g}) and integration
    by parts.
    Putting equations (\ref{tail1}) and (\ref{tail2}) together we
    finally get the leading asymptotic behavior near timelike
    infinity
 \begin{equation}\label{tail}
     f_3(t,r) \sim c\, r\, t^{-4}\,, \qquad c=-8
   \int\limits_{-\infty}^{+\infty} a(u) a'(u)^2 du\,.
 \end{equation}
 This is our main result. We claim that
 the expression
 (\ref{tail})
provides a very good approximation of the tail for solutions having
sufficiently  small initial data. More precisely, we conjecture that
for any given smooth compactly supported  functions $\alpha(r)$ and
$\beta(r)$ one can choose $\varepsilon$ such that for each fixed
$r>0$ and $t\rightarrow\infty$ the remainder $|t^4
f(t,r)-\varepsilon^3 c\, r|$ is as small as one pleases. The
numerical evidence supporting this conjecture is shown in Fig.~2.
The obvious issue remains as to whether the
  perturbation expansion corresponding to given initial data is
  convergent for sufficiently small values of $\varepsilon$.
  Without a proof of convergence our analysis is not mathematically rigorous,
  nevertheless
  we believe that it gives a rather convincing and, most importantly,
  \emph{quantitative} description of the late-time
  tail.

\begin{figure}[tbh]
\centering
\includegraphics[width=0.65\textwidth]{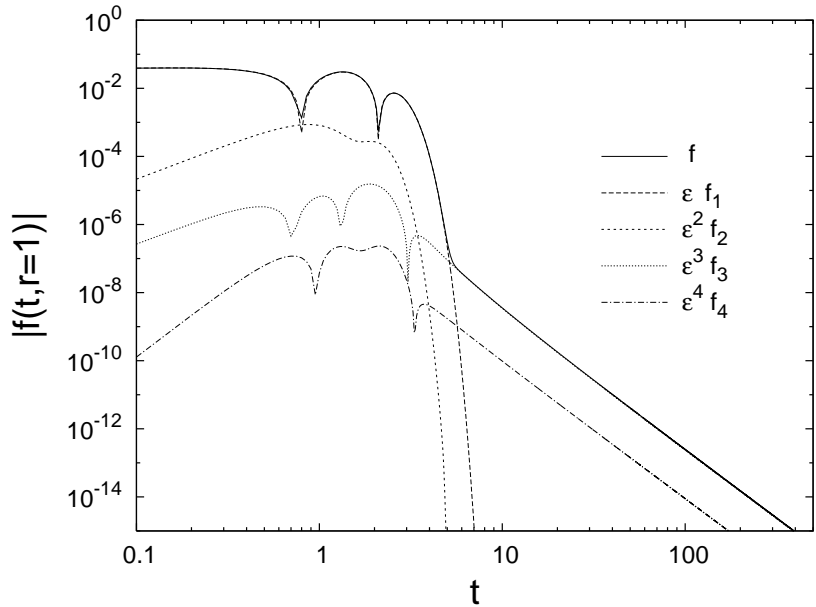}
\caption{\small{We plot (on log-log scale) the numerical solution
$f(t,r=1)$ of the initial value problem (\ref{ym})-(\ref{id}) for
$\alpha(r)=r \exp(-r^2), \,\beta(r)=r(2-r^2)\exp(-r^2)$ and
$\varepsilon=0.1$, and compare it with the third-order approximation
$\varepsilon f_1+\varepsilon^2 f_2+\varepsilon^3 f_3$ (produced by
solving numerically the perturbation equations (6)-(8)). These two
functions are indistinguishable at the scale of the figure so both
are depicted by the single solid line. The contributions of the
individual iterates are superimposed to demonstrate that the tail
comes from $f_3$. The fourth-order iterate $f_4$ serves as an
estimation of the error. Note that $f_4$ has the same late-time
slope (decay rate) as $f_3$, in agreement with Remark~2. The fit of
the function $C t^{-\gamma} \exp(A/t)$ to the full solution $f(t,1)$
for late times gives $\gamma \simeq 4.001$ and $C\simeq
2.391\!\cdot\! 10^{-5}$ which is ca. $\!4\%$ off the third-order
prediction $\varepsilon^3 c=\varepsilon^3 155
\sqrt{3\pi}/20736\approx 2.295\!\cdot\! 10^{-5}$ obtained by
evaluating the integral in (\ref{tail}).} }\label{fig2}
\end{figure}
\vskip 0.2cm A few remarks are in order:
 \vskip 0.2cm \noindent \emph{Remark~1:}
It should be clear from the above derivation that the simplicity of
the final result (\ref{tail}) is due to some amazing cancelations
(notably those occurring in equations (\ref{f2as}) and
(\ref{tail2})) which in turn are attributed to the particular form
of the nonlinearity of the Yang-Mills equations. In this respect the
Yang-Mills equations are exceptional and particularly interesting
mathematically; for most other nonlinear perturbations of the wave
equation the tail is a second-order phenomenon which is much easier
to analyze (e.g., see \cite{bcr}).

\vskip 0.2cm \noindent \emph{Remark~2:} Note that all iterates $f_k$
behave as $\mathcal{O}[(v-u)^{-1}]$ near null infinity and therefore
at each order of the perturbation expansion the sources  behave as
$\mathcal{O}[(v-u)^{-3}]$. For such sources one might expect by
dimensional arguments that there would be a $t^{-3}$ tail.
Fortunately, due to the identity
\begin{equation}\label{higher}
    \int\limits_{t-r}^{t+r} \frac{(v-t)(t-u)+r^2}{(v-u)^3}
    dv = 0
\end{equation}
all coefficients of hypothetical  $t^{-3}$ tails vanish identically
and thus all higher-order terms in the perturbation expansion decay
 as $t^{-4}$. This fact is crucial since otherwise the third-order
approximation would break down for late times; for instance a
nonzero fourth-order term $\sim \varepsilon^4/t^3$ would make the
formula (\ref{tail}) useless for times $t\gtrsim 1/\varepsilon$.
 \vskip 0.2cm \noindent \emph{Remark~3:} Note that equation
 (\ref{ym}) has the scaling symmetry: if $f(t,r)$ is a solution, so
 is $f_{\lambda}(t,r):=\lambda f(\lambda t,\lambda r)$. Under this
 scaling the energy scales as $E(f_{\lambda})=\lambda E(f)$,
 hence given any finite energy initial datum one can scale it down
  to an arbitrarily small amplitude and energy.  Note, however, that for compactly supported initial data
   such a rescaling
  spreads the support by a factor $1/\lambda$ and  for this reason it
  cannot make
  large data smaller in the sense of our perturbation expansion.
   This follows immediately from the fact that all iterates
  $f_k(t,r)$ in (\ref{pert}) scale in the same way. In other words, the rescaling
  does not change the convergence properties of the perturbation
  expansion.
\section{Schwarzschild background} On the exterior Schwarzschild
spacetime
\begin{equation}\label{schw}
    ds^2=-\left(1-\frac{2m}{r}\right) dt^2 + \left(1-\frac{2m}{r}\right)^{-1} dr^2 + r^2
    (d\theta^2 + \sin^2\theta d\phi^2)\,, \qquad r > 2m\,,
\end{equation}
the spherically symmetric Yang-Mills equation corresponding to the
ansatz (\ref{ansatz})  takes the form
\begin{equation}\label{eqwsch}
  \left(1-\frac{2m}{r}\right)^{-1}  \ddot w -
  \left(\left(1-\frac{2m}{r}\right)w'\right)'-\frac{1}{r^2} w
  (1-w^2)=0\,.
\end{equation}
When $m=0$ this equation  reduces of course to (\ref{eqw}). In terms
of the new variables
\begin{equation}\label{var}
x=r+2m \ln\left(\frac{r}{2m}-1\right)\,, \qquad
h(t,x)=w(t,r(x))-1\,,
\end{equation}
equation (\ref{eqwsch}) becomes
\begin{equation}\label{eqwx}
    \ddot h -\frac{d^2h}{dx^2} +\left(1-\frac{2m}{r}\right)\frac{2}{r^2}
    h=
  -\left(1-\frac{2m}{r}\right)\frac{1}{r^2}(3h^2+h^3)\,,
\end{equation}
where $r=r(x)$. Dropping the nonlinear terms on the right side of
(\ref{eqwx}) one gets the linear $1+1$ dimensional wave equation on
the real axis $-\infty <x<\infty$ with the effective potential
$V(x)=2/r^2-4m/r^3$. This equation describes the propagation of the
dipole ($l=1$) electromagnetic  perturbation of the Schwarzschild
black hole. For \emph{intermediate} times the linearized
approximation is very good; this stage of evolution has the form of
exponentially damped oscillations dominated by the fundamental
(i.e., least damped) quasinormal mode. We recall that quasinormal
modes are solutions of the linearized equation satisfying the
outgoing wave conditions $h(t,x)\sim e^{-ik(t\mp x)}$ for
$x\rightarrow \pm \infty$. In the case at hand the fundamental
quasinormal mode has the eigenvalue $k=0.49653-0.18498 i$ (in units
where $2m=1$) \cite{l}.

The quasinormal mode decays exponentially so for late times it
becomes negligible and eventually a polynomial tail is uncovered.
Since the pioneering work of Price \cite{p} it has been known that
the tail of the $l$-th multipole decays as $t^{-2l-3}$, thus for the
dipole the linearized theory predicts the tail $t^{-5}$ and this is
exactly the result derived in \cite{cw}.
We claim that this prediction is incorrect and the actual tail
behaves in the same manner as in Minkowski spacetime, namely it
decays as $t^{-4}$. Regarding  equation (\ref{eqwsch}) as the
perturbation (for $r\gg 2m$) of equation (\ref{eqw}), one can see
from dimensional considerations that the failure of linearization is
due to the fact that the linear terms in (\ref{eqwsch})
corresponding to nonzero curvature (proportional to $m$) are of
\emph{shorter range} (using PDE jargon) than the nonlinear terms.
Thus, the presence of the black hole should not alter the flat space
tail $t^{-4}$. The numerical substantiation  of this handwaving
argument is shown in Fig.~3.
\begin{figure}[tbh]
\centering
\includegraphics[width=0.55\textwidth]{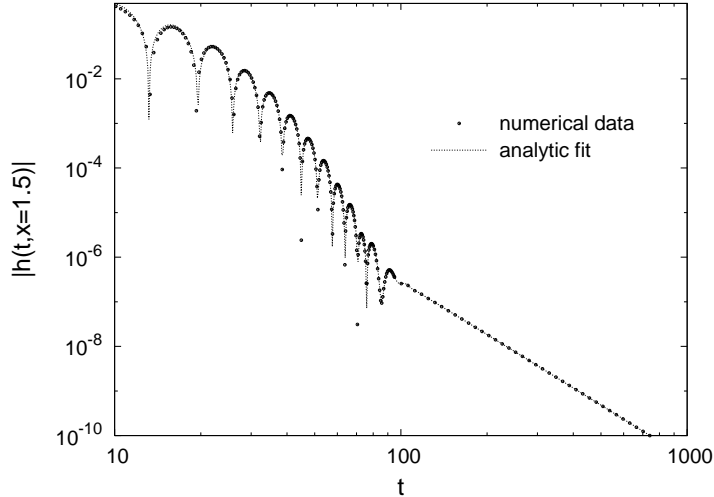}
\caption{\small{Scattering of the Yang-Mills wave off the
Schwarzschild black hole (with $2m=1$). We plot (on log-log scale)
the numerical solution $h(t,x=1.5)$ of equation (\ref{eqwx}) for the
initial data of the form of the "ingoing" gaussian $h(0,x)=A
\exp(-(x-x_0)^2/s^2)$ with $A=0.85, x_0=3, s=1.5$. Fitting the
exponentially damped sinusoid $Q(t)=B e^{-\Gamma t} \sin(\Omega
t+\delta)$ on the interval $30<t<60$ we get $\Gamma=0.184$ and
$\Omega=0.495$, in perfect agreement with the known parameters of
the fundamental quasinormal mode. The fit of the power law decay
$P(t)=C t^{-\gamma} \exp(D/t+E/t^2)$ for times $t>300$ gives
$\gamma=3.9997$. The sum $|Q(t)+P(t)|$ (depicted by the dashed line)
provides a remarkably good approximation of the full solution for
\emph{all} $t\gtrsim 20$.~It should be pointed out, however, that
our initial data were tuned a bit to maximize the effect of the
nonlinearity. If the subdominant $t^{-5}$ tail coming from the
potential has a large coefficient, i.e. the tail behaves as $C
t^{-4} + \tilde C t^{-5}$ with $C\ll \tilde C$,  then one has to
wait for a long time before the true asymptotic behavior sets in
(which might be misleading without an analytic insight).}
}\label{fig3}
\end{figure}

Unfortunately, for the Schwarzschild background we were not able to
derive a quantitative formula,  analogous to (\ref{tail}), relating
the amplitude of the tail to initial data. An attempt to repeat the
perturbation analysis from section~II  encounters serious
difficulties on Schwarzschild background which are caused by the
violation of Huygens' principle in $1+1$ dimensions and the presence
of the potential. It would be interesting to pursue this problem
further, perhaps borrowing ideas from an approach proposed some time
ago by Barack \cite{b}. Although Barack considered only the linear
wave equation, we wish to emphasize that there are many similarities
between his work and our analysis in section~II.

\section{Conclusions} Using third order nonlinear perturbation theory we
determined the late-time tail of spherically symmetric Yang-Mills
equations on Minkowski background. We also gave heuristic arguments
that the same tail is present on Schwarzschild background. In both
cases we provided numerical evidence supporting our results. We hope
that our approach will trigger more rigorous mathematical analyses
of these physically important phenomena.

 We remark that the ideas presented here
can be applied to other nonlinear wave equations. For example, one
can show by similar methods that for the semilinear wave equation
$g^{\mu \nu}\nabla_{\mu}\nabla_{\nu} \phi + |\phi|^{p}=0$ on
Minkowski background the tail decays as $t^{1-p}$ for
$p>1+\sqrt{2}$, while on the Schwarzschild background the tail
changes its character at $p=4$ from linear (Price's law $\phi \sim
t^{-3}$ for $p\geq 4$ \cite{dr}) to nonlinear ($\phi \sim t^{1-p}$
for $1+\sqrt{2}<p<4$). A systematic analysis of the competition
between linear and nonlinear effects in scattering for semilinear
wave equations in Minkowski spacetime will be given elsewhere
\cite{bcrs}. \vskip 0.3cm \noindent \textbf{Acknowledgments:} P.B.
thanks Bernd Schmidt, Helmut Friedrich, Alan Rendall, and Nikodem
Szpak for helpful discussions. This research was supported in part
by the Polish Research Committee grant 1PO3B01229.
\noindent \subsection*{Appendix} We derive here the equation
(\ref{f2nul}). Our starting point is the equation (\ref{eqf2}) in
which we relabel coordinates $(u,v)\rightarrow (u',v')$ and use the
retarded time $u=t-r$:
\begin{equation}\label{f2a}
    f_2(u,r)=-\frac{3}{4r}
    \int\limits_{|u|}^{u+2r} dv'  \int\limits_{-v'}^{u}
    \frac{u'+v'-2u+(uu'+uv'-u'v'-u^2)/r}{v'-u'} f_1^2(u',v')
    du'\,.
\end{equation}
We let $\epsilon=1/r$ and define the quantity
\begin{equation}\label{i}
    I(u,\epsilon):=-\frac{3}{2}
    \int\limits_{|u|}^{u+2/\epsilon} dv'  \int\limits_{-v'}^{u}
    \frac{u'+v'-2u+\epsilon(uu'+uv'-u'v'-u^2)}{v'-u'} f_1^2(u',v')
    du'\,.
\end{equation}
Expanding this in Taylor's series
    $I(u,\epsilon)=A(u)+B(u)\epsilon+ \mathcal{O}({\epsilon}^2)$ we obtain
\begin{eqnarray}
   A(u)&:=& I(u,0)=-\frac{3}{2}
    \int\limits_{|u|}^{\infty} dv'  \int\limits_{-v'}^{u}
    \frac{u'+v'-2u}{v'-u'} f_1^2(u',v')
    du'\,,\\
B(u)&:=&\frac{\partial
I(u,\epsilon)}{\partial\epsilon}\Bigr\rvert_{\epsilon=0}:=
g(u)+h(u)\,,
\end{eqnarray}
where
\begin{eqnarray}
   h(u)&=& -\frac{3}{2}
    \int\limits_{|u|}^{\infty} dv'  \int\limits_{-v'}^{u}
    \frac{uu'+uv'-u'v'-u^2}{v'-u'} f_1^2(u',v')
    du'\, \label{h} \\
 g(u)&=& 3 \int\limits_{-\infty}^{u} a'(u')^2
    du'\,.   \label{g}
\end{eqnarray}
An elementary calculus exercise shows that
\begin{equation}
    h'(u)=A(u)\,.
    \end{equation}
Putting all the above equations together and noting that $r=(v-u)/2$
we get equation (\ref{f2nul}).

\end{document}